\documentclass{iaus}
\usepackage{graphicx}

\title[Stars with large proper motions] 
{A New All-Sky Satalog of Stars with Large Proper Motions}

\author[S. L\'epine {\it et al.}]
{S\'ebastien L\'epine$^1$, Michael M. Shara$^1$, R. Michael Rich$^2$,
  A. Wittenberg$^1$, M. Halmo$^1$, \and B. Bongiorno$^1$}

\affiliation{$^1$Dept. of Astrophysics, American Museum of
  Natural History, Central Park West at 79th street, New York, NY
  10024, USA \\ email: {\tt lepine@amnh.org, mshara@amnh.org}
  \\[\affilskip]
  $^2$Dept. of Astrophysics, University of California at Los Angeles
  {\tt rmr@astro.ucla.edu}}

\pubyear{2008}
\volume{248}  
\pagerange{??--??}
\jname{A Giant Step: From Milli- To Micro- Arcsecond Astrometry}
\editors{A.C. Editor, B.D. Editor \& C.E. Editor, eds.}
\begin{document}

\maketitle


\begin{abstract}
A new all-sky catalog of stars with proper motions $\mu>0.15''$
yr$^{-1}$ is presented. The catalog is largely a product of the
SUPERBLINK survey, a data-mining initiative in which the entire
Digitized Sky Surveys are searched for moving stellar
sources. Findings from earlier proper motions surveys are also
incorporated. The new all-sky catalog supersedes the great historic
proper motion catalogs assembled by W. J. Luyten (LHS, NLTT), and
provides a virtually complete $>98\%$ census of high proper motion
stars down to magnitude $R=19$.

\keywords{Catalogs, solar neighborhood, stars: kinematics, stars:
  low-mass, brown dwarfs, stars: subdwarfs, stars: white dwarfs,
  Galaxy: stellar content}
\end{abstract}


\firstsection 
\section{The SUPERBLINK survey}

We have been conducting an all-sky survey for stars with large proper
motions using data from the Digitized Sky Surveys (DSS). The scanned
images in the DSS cover the entire sky in multiple bands and at
various epochs, and the temporal baseline between the earliest and
latest epoch is between 15 and 45 years for most areas on the sky. 

The large motion displayed by high proper motion stars between the two
epochs is detected directly from the scans by means of an image
subtraction algorithm, described in detail in \cite[L\'epine \etal
(2002)]{LSR02}. Two-epochs finder charts are generated, which can be
blinked on the computer screen. All objects detected in the survey are
thus verified by eye, and spurious detections are excluded.

At the bright end, stars tend to become saturated on the DSS images,
and are no longer properly detected by the code. The TYCHO-2 catalog
is used to complete the census at the bright end. Also, all stars
detected by SUPERBLINK are searched for a counterpart in the
TYCHO-2. The positional and proper motion information from the TYCHO-2
catalog are used for all matching counterparts. The two-epoch charts
are however examined to verify consistency; in some cases, it is found
that the proper motion from TYCHO-2 must be in error, and the
SUPERBLINK proper motion is used instead.

Counterparts in the the 2MASS All-Sky Catalog of Point
Sources \cite[Cutri \etal (2003)]{C03} are also identified for all
SUPERBLINK detections. At the faint end, the positions are thus those
extrapolated from the 2MASS catalog, and are thus realized in the ICRS
system and accurate to about 0.1''. Counterpart are also identified in
the USNO-B1 catalog \cite[Monet \etal (2003)]{M03}, and together with
2MASS, provide optical and infrared magnitudes for almost all the
stars.

The systematic comparison with the 2MASS catalog allows allows us to
systematically identify all common proper motion doubles which are
resolved in the 2MASS images. The CCD observations from 2MASS have
significantly higher resolution than the photographic images from the
DSS. Numerous common proper motion doubles and multiple systems have
being identified and their components will be listed as separate
entries in the LSPM catalog. Many faint companions of Hipparcos stars
are also being identified this way \cite[e.g. L\'epine \& Bongiorno
(2007)]{LB07}.

\begin{figure}[t]
\begin{center}
 \includegraphics[width=5in]{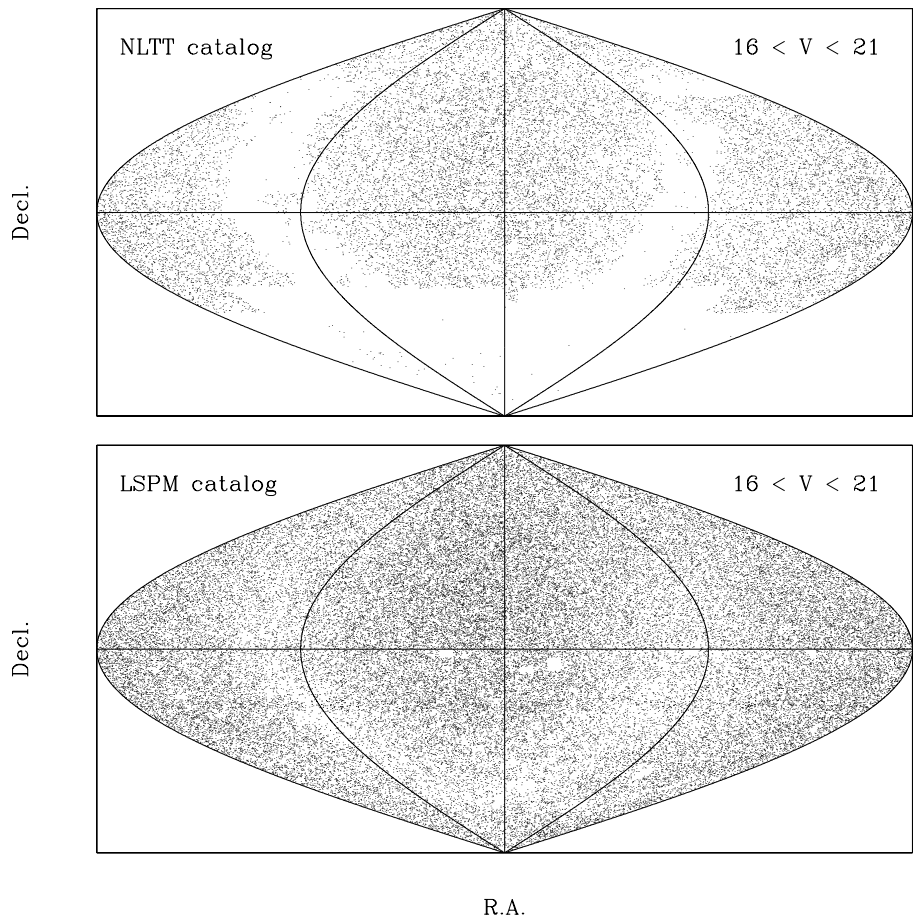} 
 \caption{Comparison between the distribution of faint ($V>16$) stars
 from the NLTT catalog (top) and the new LSPM catalog (bottom). The
 new catalog fills in all the gaps of the NLTT, particularly in the
 low Galactic latitude fields, and provides the most complete all-sky
 census of high proper motion stars to date.}
 \label{fig1}
\end{center}
\end{figure}

\section{Replacing the NLTT catalog}

The NLTT catalog \cite[Luyten (1979)]{L79} was notoriously incomplete
in two main regions: the sky south of Decl.=-30$^{\circ}$, and areas
of high stellar density along the plane of the Milky Way. The
incompleteness was most severe for stars fainter than magnitude V=16
(see Fig.1). 

The northern part of the SUPERBLINK survey was completed first, and
the results have already been published in \cite[L\'epine \&
  Shara(2005)]{LSR05} as the LSPM-north catalog. The full LSPM catalog
now fills in most of the remaining gaps in the south, and at last
provides a true, all-sky census of faint stars with large proper
motions (Fig.1). While there remains some level of incompleteness at
low Galactic latitudes, especially toward the Galactic center, most of
the variations in surface density observed in Fig.1 are due to
selection effects from the high proper motion cutoff ($\mu>0.15''$
yr$^{-1}$) of the survey. A combination of the Sun's motion through
the local standard of rest and the asymmetric drift of the Galactic
thick disk and halo stars results in more stars having large
transverse motions at high Galactic latitudes, hence the large
density of stars detected there.

Compared with the 58,845 stars with proper motions $\mu>0.18''$
yr$^{-1}$ listed in the NLTT catalog, the LSPM catalog will list over 
122,000 stars with proper motions $\mu>0.18''$ yr$^{-1}$. With the
increased sky coverage and completeness, the LSPM catalog makes the
Luyten catalog obsolete, and from now on should be used as a
replacement to the NLTT. For convenience, all NLTT stars will also be
identified in the LSPM both by their LHS designation and NLTT catalog
number.

\section{Stellar contents and kinematics}

A reduced proper motion diagram shows the stars in the LSPM to be of
three main classes. Low-mass K and M red dwarfs from the
disk population dominate, but a significant fraction are low-mass
subdwarfs from the halo (sdK, sdM), and the catalog also contains
thousands of white dwarfs. While we currently lack parallax distances
for most of the stars in the LSPM catalog, photometric distances can
be calculated for specific classes of objects. The red dwarfs, in
particular, have a reasonably well calibrated $[M_v,V-J]$
color-magnitude relationship \cite[L\'epine (2005)]{L05}. 

With photometric distances and proper motions, it is possible to
investigate the local kinematics of the red dwarfs in the vicinity of
the Sun ($d<100$pc). By selecting stars is specific parts of the sky,
one can obtain velocity-space projections in the UV, UW, and VW plane
(Figure 2). Because of the high proper motion cutoff of the LSPM
catalog ($\mu>0.15''$ yr$^{-1}$, stars with low projected velocities
are not represented in the census, which leaves a low-velocity
``hole'' in the maps of projected velocities. The hole increases for
stars at larger distances. Despite this artifact, one can see that the
velocity space projections of the nearby red dwarfs are not isotropic
and show considerable structure. A comparison with the velocity space
distribution of Hipparcos stars calculated by \cite[Nordstr\"om \etal
(2004)]{N04}, shows very good agreement with our data. This show how
future astrometry, providing accurate parallaxes for all the LSPM
stars, may have a major impact in uncovering fine structure in the
kinematics of stars in the Solar vicinity.

\begin{figure}[t]
\begin{center}
 \includegraphics[width=6in]{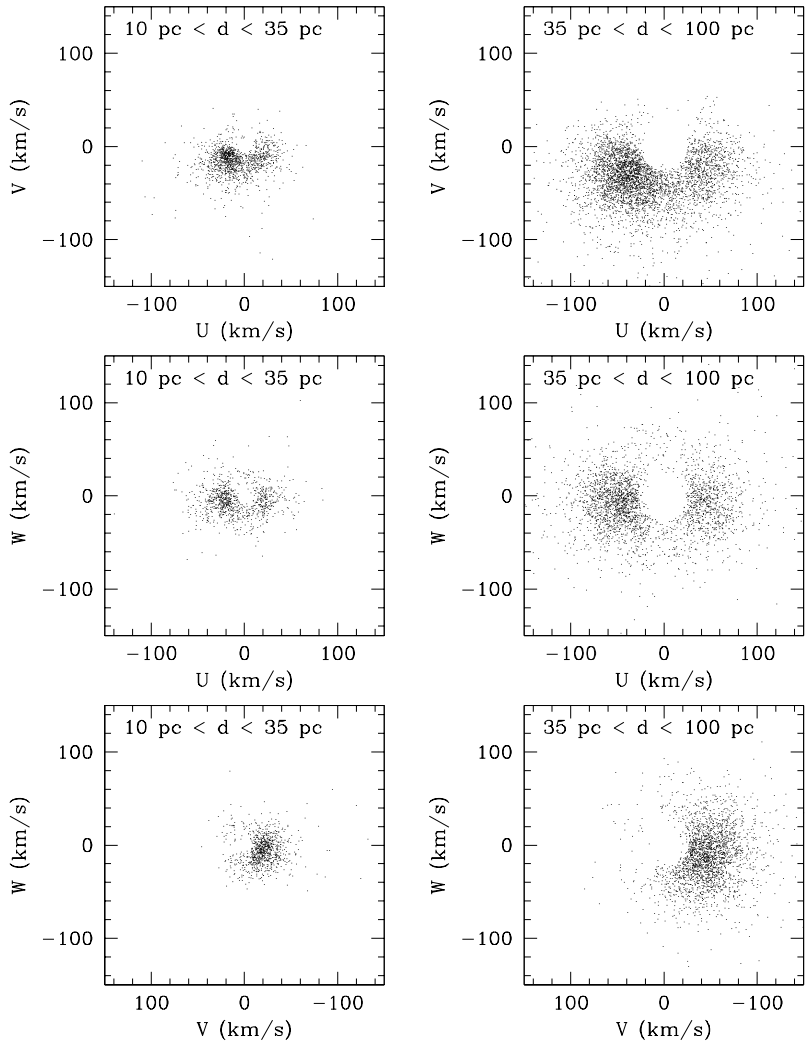} 
 \caption{Projected motions in the UVW of red dwarf stars in the Solar
   neighborhood, based on LSPM catalog proper motions and photometric
   distances.The UW projection is from stars found at low Galactic
   latitude in the direction of the apex and antapex of the Sun's
   orbital motion around the Galaxy. The VW projection is obtained
   from stars in the direction of the Galactic center and
   anti-center.}
 \label{fig2}
\end{center}
\end{figure}

\section{Conclusions}

The LSPM catalog now has all-sky coverage. The LSPM-south catalog will
complement the already released LSPM-north, and yield a highly
complete catalog of stars with proper motion $\mu>0.15 ''$
yr$^{-1}$. The catalog is estimated to be $>98\%$ complete for all
H-burning stars and white dwarfs with proper motions in the range
above, covering virtually all objects down to visual magnitude
19. The catalog is realized at the bright end by the Tycho-2 catalog,
down to magnitude $\approx10-12$. At the faint end, which encompass
the vast majority of the stars, the proper motions are obtained from
the SUPERBLINK software, while the positions are determined by the
counterparts in the 2MASS catalog. Overall, the positional accuracy of
the catalog is thus better than $0.12''$, while proper motions at
the faint end have typical errors $\approx10$ mas yr$^{-1}$ north of
Decl.=-30$^{\circ}$, and $\approx20$ mas yr$^{-1}$ south of this. All
double stars which are resolved in the 2MASS survey have been
identified and are listed individually.

The SUPERBLINK survey is now being expanded to lower proper motion
regimes, and future releases will expand the catalog to proper motions
$\mu>40$ mas yr$^{-1}$.

\begin{discussion}
\end{discussion}

\end{document}